\documentclass[smallextended]{svjour3}       
\smartqed  
\usepackage{graphicx}
\begin{document}

\title{ Gravitation Astrometric Measurement Experiment }

\author{Mario Gai \and
       Alberto Vecchiato \and
       Sebastiano Ligori \and 
       Alessandro Sozzetti \and 
       Mario G. Lattanzi
}
\authorrunning{ Gai, Vecchiato, Ligori, Sozzetti and Lattanzi } 

\institute{M. Gai \at
Istituto Nazionale di Astrofisica - Astronomical Observatory of Torino \\
              Tel.: +39-011-8101943; 
              Fax: +39-011-8101930; 
              \email{gai@oato.inaf.it}           
}

\date{Received: date / Accepted: date}

\maketitle

\begin{abstract}
The Gravitation Astrometric Measurement Experiment (GAME) is a mission 
concept based on astronomical techniques (astrometry and coronagraphy) 
for Fundamental Physics measurements, namely the $\gamma$ and $\beta$ 
parameters of the Parametrized Post-Newtonian formulation of gravitation 
theories extending the General Relativity. 
The science case also addresses cosmology, extra-solar planets, 
Solar system objects and fundamental stellar parameters. 
The mission concept is described, including the measurement approach and the
instrument design.

\end{abstract}

\section{Introduction}
\label{intro}
The experiment of Dyson, Eddington and Davidson, whose concept is sketched in
the left panel of Fig.~\ref{fig:Eddington}, gave the first confirmation of
Einstein's General Relativity theory by observations of known stellar fields during the May 29th, 1919 eclipse. It
measured the apparent positions of a few stars, within a few degrees from the
solar limb during the eclipse, compared to their unperturbed relative positions
(e.g. in night time observations a few months away). 
The arc variation is interpreted in terms of light deflection, providing 
an estimate of the related $\gamma$ parameter of the Parametrised Post-Newtonian 
formalism (PPN, Sec.~\ref{subsec:gamma}) with precision $\sim10\% =10^{-1}$. 
Measurements of light deflection from
the ground are affected by several shortcomings, like to the short eclipse
duration, the high background flux from the solar corona, the atmospheric
disturbances and the limited number of bright sources in the field related to a
given eclipse, which undermine their competitiveness because of fundamental
reasons and cannot be relieved significantly by technological improvements. 
The current best estimate of $\gamma$ from deflection measurements at radio 
wavelength is in the range $10^{-4}$ with VLBI \cite{2004PhRvL..92l1101S}, 
or $10^{-5}$ from the Cassini experiment \cite{2003Natur.425..374B}. 

The $\beta$ parameter, instead, is usually estimated through accurate orbit
determination of Solar System bodies which put to evidence the effect of
perihelion shift excess. This effect is larger for orbits close to the Sun and
with high eccentricities, the most convenient targets being planet Mercury and
NEOs, which can be more conveniently observed from space.
The so called ``grand-fits'' of the Solar System ephemeris were able to reach 
the $10^{-4}$ level thanks to their long temporal range (1963-2003) and 
the inclusion of other data coming from radar observations of inner bodies 
\cite{2005AstL...31..340P,2010IAUS..261..170P}. 

The main scientific objective of GAME is the estimation of such so-called 
Eddington parameters, crucial to the PPN formalism, 
which is commonly used to test the metric theories of gravity in the 
weak-field regime of the Solar System. 
\emph{The goal precision is a few $10^{-8}$ on $\gamma$, and 
a few $10^{-6}$ on $\beta$. } 
\\ 
The test of  gravity theories to this level of accuracy has significant 
implications for our understanding of several physical and astrophysical 
issues at a very fundamental level. 
In particular, although we refer to the performance on PPN parameters 
for ease of comparison with previous experiments, the measurement 
interpretation at this level is probably best set in the 
Parametrised Post-Post-Newtonian (PPPN) framework. 

GAME can measure the light deflection effects associated to oblate
and moving giant planets (Jupiter, Saturn), never detected yet. 
\\ 
Also, GAME can provide significant results in the field of extra-solar 
planetary systems: as a pointed astrometric mission in the period 2020-2025, 
it can be used  for long term follow-up of candidates identified by Gaia, 
Kepler and Corot, providing complementary information on heavy companions 
at a few AU from the parent star thanks to its long temporal baseline 
(about 15 years) with respect to the detection. 
\\ 
During its normal operation, but over a small fraction of the mission 
lifetime, GAME may perform many observations of solar system objects. 
Lunar and planetary occultations can provide stellar radii and binary 
separations over a sample of up to several thousand objects, and a 
relevant contribution will be provided on Solar System dynamics, in 
particular on Near Earth Object orbit determination. 
Another scientific goal can be identified in the high angular resolution 
monitoring of the Corona and circumsolar environment.

The GAME concept was previously presented assuming a smaller 
implementation scale, and slightly different technical aspects, 
as discussed in \cite{2009AdSpR..44..579V} and references therein. 

\section{Scientific objectives and requirements}
\label{sec:sciobj}
\subsection{Estimation of the PPN $\gamma$ parameter}
\label{subsec:gamma}
The PPN formalism \cite{2006lrevr...9..3..W} was
introduced to classify all of the conceivable metric theories of gravity at
their Post-Newtonian level by means of a given set of parameters. Each theory is
characterized by the values taken by these parameters, which are also associated
to specific physical meanings. In this framework, the $\gamma$ parameter
quantifies the effect of mass on space-time curvature, while $\beta$ is related
to the non-linearity of the superposition of the gravity fields of different
bodies, and both are unity in General Relativity (GR). However, General
Relativity can act as a cosmological attractor for scalar-tensor theories of
gravity with expected deviations in the $10^{-5}-10^{-7}$ range for $\gamma$
\cite{1993PhRvL..70.2217D}. Also, during the last decade, a strong
experimental evidence of an acceleration of the expansion of the Universe at the
present time has been provided by several observational data. This has been
interpreted in the concordance $\Lambda$CDM scenario as the effect of a long
range perturbation of the gravity field of the visible matter generated by the
so-called Dark Energy. These data add to those available for long time at
different scale length, which are explained with the existence of non-barionic
Dark Matter (e.g. galaxy rotation curves) or with some kind of modification of
the General Relativity theory (e.g. Pioneer anomalies
\cite{2008PhRvD..78j3001B,2008APS..APR.S1010L,2009SSRv..148..169T}. However,
there are claims that these data could also be explained with a modified version
of General Relativity, in which the curvature invariant $R$ is no longer
constant in the Einstein equations ($f(R)$ gravity theories). 
Present experimental data are not accurate enough to discriminate between 
these scenarios, but this could be done with a $10^{-7}$ level, or better, 
measurement of $\gamma$
\cite{2008JCAP...03..024B,2005PhRvD..72d4022C,2009MNRAS.394..947C,2010PhLB..686...79C}.

From a phenomenological point of view, the $\gamma$ parameter is associated to
the light deflection, and it is easy to demonstrate \cite{2009AdSpR..44..579V} 
that the accuracy estimation of the $\gamma$ parameter is proportional to that
of the light deflection, which is directly related to the precision of the
angular separation measurements.

\begin{figure}
  \begin{center}
    \includegraphics[width=0.48\textwidth]{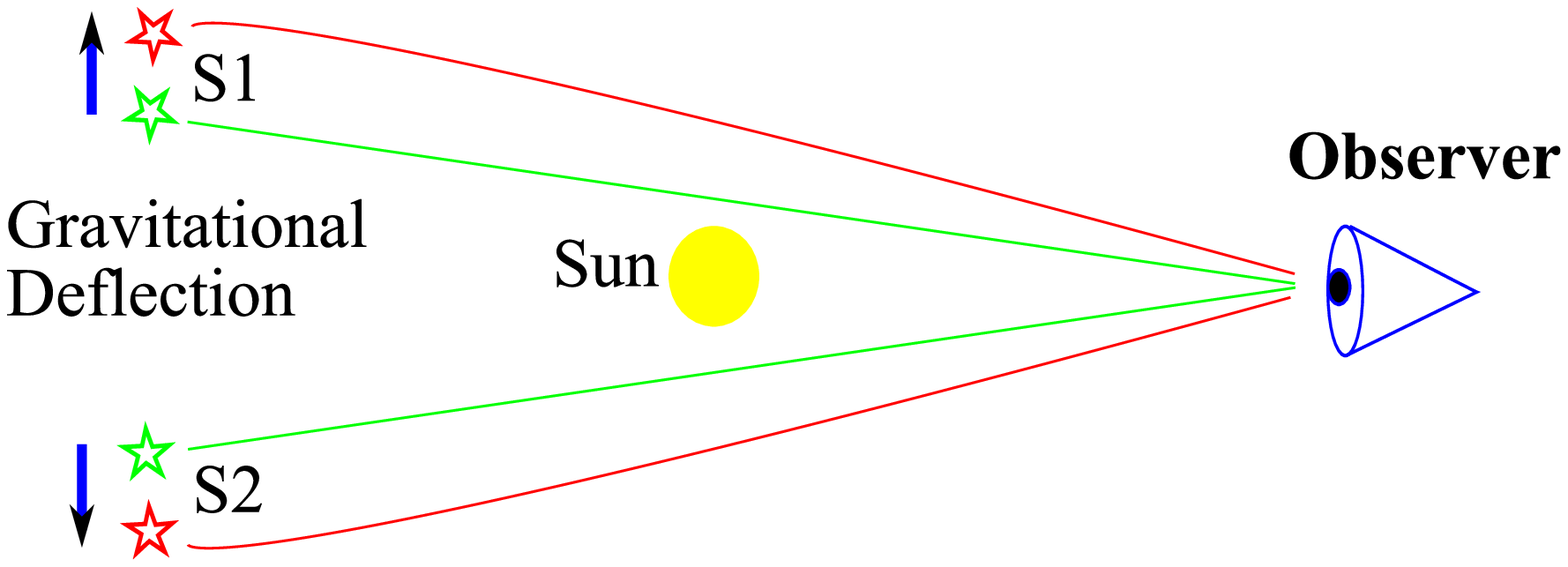}
    \includegraphics[width=0.48\textwidth]{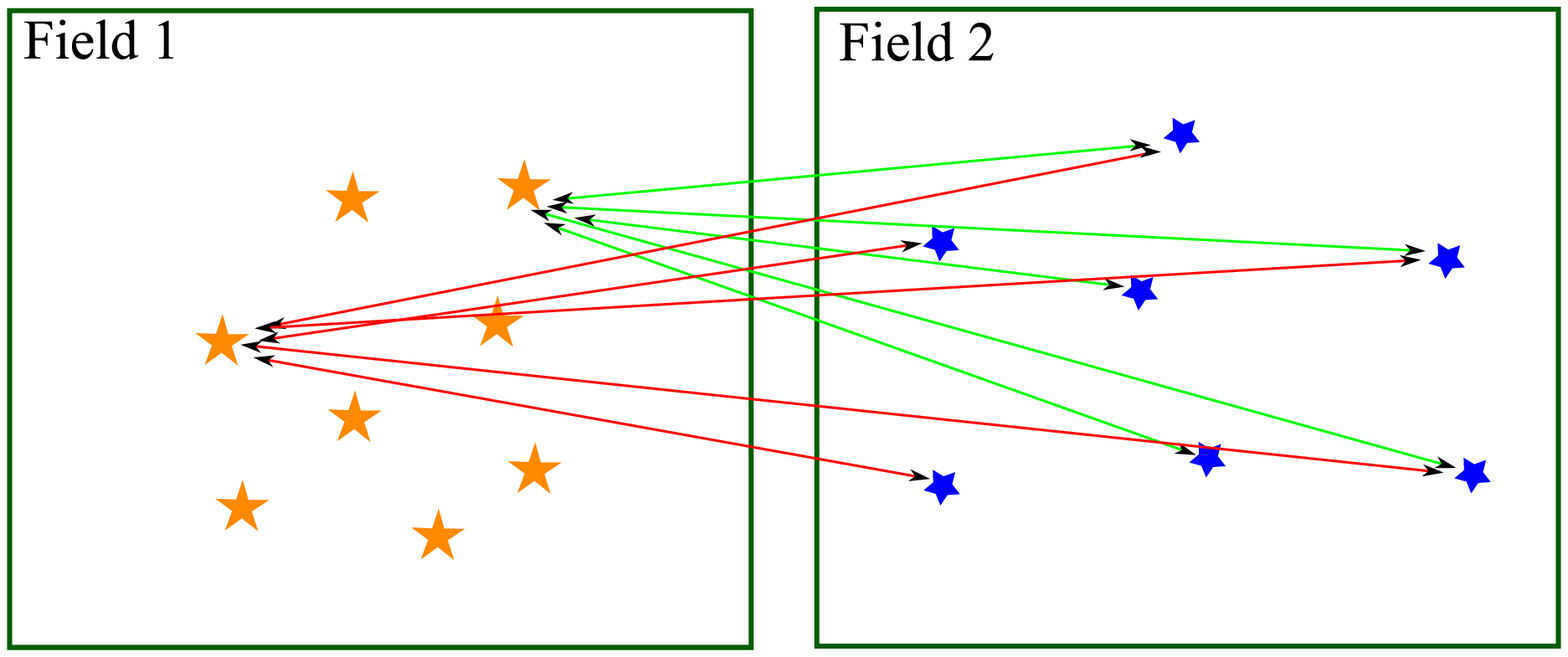}
  \end{center}
\caption{Representation of light deflection by the Solar gravitational field on
the light rays from stars. The apparent star position is displaced from the Sun
because of the photon path bending (left). On the right, the observable: arcs
between stars from each field, superposed on the detector. Arc length changes
with time, depending on pointing position vs. the Sun.}
\label{fig:Eddington}
\end{figure}

\begin{figure}
  \begin{center}
    \includegraphics[width=0.70\textwidth]{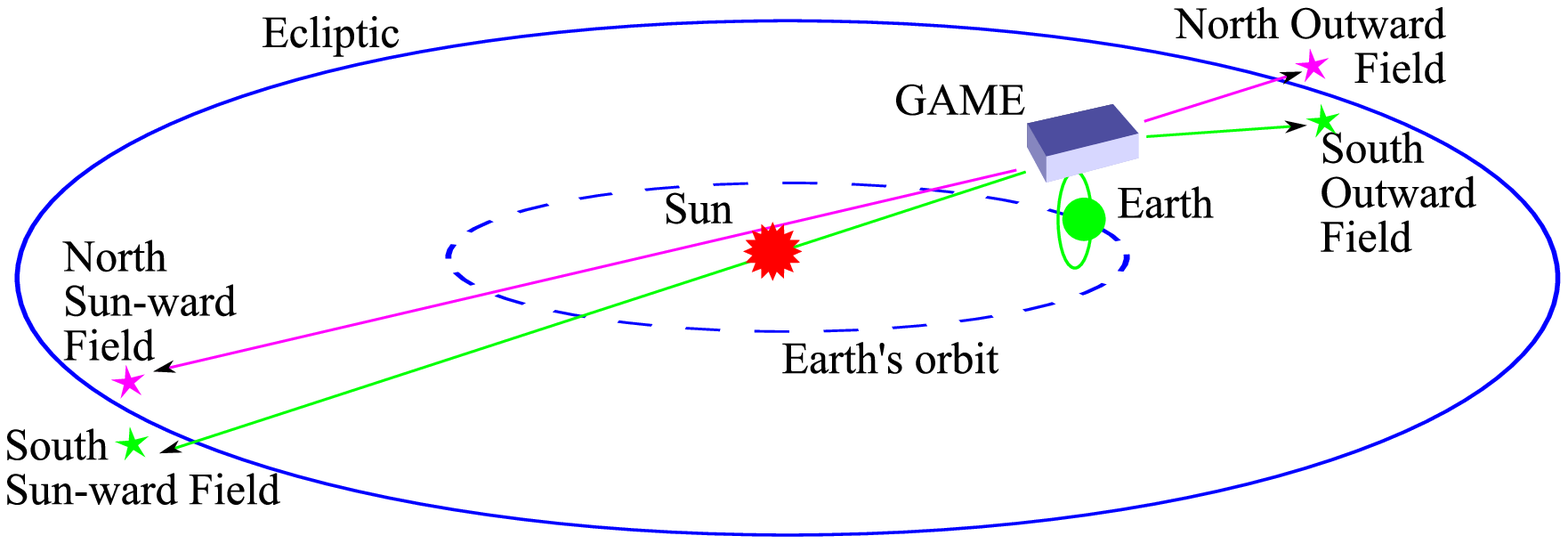}\hfill
    \includegraphics[width=0.25\textwidth]{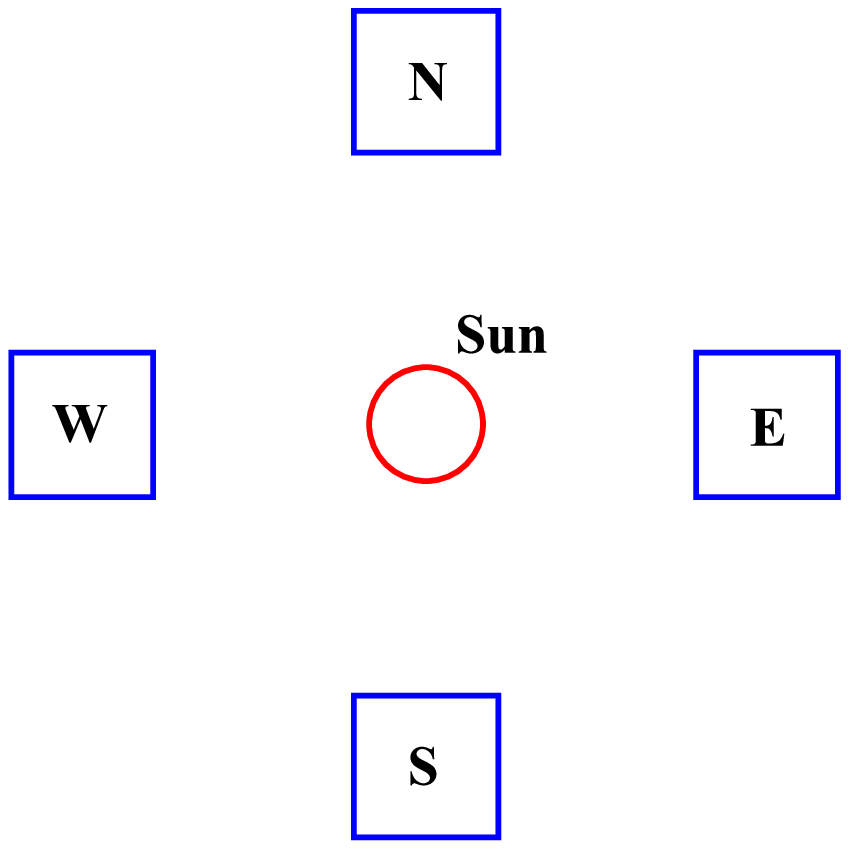}
  \end{center}
\caption{The GAME satellite orbiting the Earth and observing fields 
respectively in conjunction and opposition to the Sun (left); relative 
placement of the four Sun-ward fields (right).} 
\label{fig:ObsGeometry}
\end{figure}

Light deflection reaches a peak value of $1''\!\!.74$ at the solar limb, and
decreases rapidly at increasing angular distance, and therefore, in order to
estimate the $\gamma$ parameter at the $10^{-6}$ level and beyond, microarcsec
($\mu$as) level measurements of relative star positions (Fig.~\ref{fig:Eddington},
right) are required at a few degrees from the Sun. 
Previous simulations showed that the $10^{-7}$ level of accuracy could 
be reached within the baseline of the same measurement concept scaled down 
to fit a small mission framework \cite{2009AdSpR..44..579V}. 
In this scenario the satellite observes two (North/South oriented) $7'\times7'$ FOVs at $2^\circ$ from the Sun centre for $\sim8$ months in two years. 
Observing crowded stellar fields toward the
Galactic centre increases the performance, and the inclusion of specific bright
stars elsewhere around the ecliptic as additional targets further improves on
the final accuracy. The targeted accuracy of the medium size mission GAME
(Fig.~\ref{fig:ObsGeometry}, left) comes from improvement factors as the
\emph{longer mission duration\/}, the adoption of \emph{four}
(instead of two) \emph{larger fields of view\/} set at $90^\circ$ from each
other (Fig.~\ref{fig:ObsGeometry}, right), and a higher performance optical
configuration which also allows for a \emph{better control of systematic
errors.}

In addition to the four Sun-ward fields, GAME also observes at the same time,
and using as much as possible the same parts of the instrument, four fields with
the same relative placement in the direction opposite to the Sun, hereafter
labelled ``outward'' direction/fields.

Star location estimation at a small fraction of arcsec, for a precise
determination of the light deflection angle and therefore of $\gamma$, requires
imaging with a reasonable resolution, $\sim0''\!\!.1$, and high photometric
signal to noise ratio (SNR), achievable with a medium size telescope (diameter
$\sim1.5$~m).

The individual star coordinates in both epochs must be measured with typical
random error $\sigma_\mathrm{C}$. The collective location measurement of
$N_\mathrm{S}\simeq5\times10^5$ stars, over $N_\mathrm{E}=5$ epochs, 
is affected by a residual error 
$\sigma_\mathrm{RC} \approx \sigma_\mathrm{C}/ \sqrt{N_\mathrm{S}N_\mathrm{E}}$. 
This must be a fraction of the typical deflection difference between stars 
in opposite fields ($2\Delta\psi\simeq470$~mas), smaller than the goal of 
GAME on $\gamma$, i.e. a few $10^{-8}$, so that 
$\sigma_\mathrm{C}\approx10^{-8}\Delta\psi\times N_\mathrm{S}N_\mathrm{E} 
\simeq 10~\mu$as.

Such precision level is actually achieved on the bright end of the sample,
bearing the most significant statistical weight in the data composition. This
precision requires direct measurement also of unperturbed positions, as well as
of deflected ones, since they are not available from other sources (e.g. the
Gaia catalogue). Also, the homogeneous data set is apt to consistent calibration
and removal of systematic errors.

A detector with a field size of several square degrees, with the desired 
$\sim 0''\!\!.1$ resolution, imaging an extended region around the Sun, 
similarly to the 1919 Eddington et al. experiment, is impractical for GAME. 
Setting a field size of $30'\times30'\approx3\times10^8$ pixels, with a 
projected separation on the sky corresponding to a {\em base angle} (BA) 
of about $4^\circ$, discussed below, significantly alleviates the detector 
requirement, taking it into the complexity order e.g. of Gaia. 

Adoption of a highly symmetrical instrument design mitigates the calibration
requirements, as the instrument response is the same, at first order, for all
fields. Besides, using as far as possible the same optical paths and detector,
perturbations are expected to act mainly in common mode over the signal in
different fields, thus reducing their influence on the differential measurement.
Finally, the observation of several fields, with only some of them affected by a
significant amount of deflection, implements an efficient simultaneous
calibration, in which the payload evolution is monitored in real time.

Observation of the same fields at different epochs, with the Sun displaced by a
large angular amount from either field due to the Earth's orbital motion,
provides the ``true'' distance between the objects. The variation of star
separation between the two epochs defines the angular value of light deflection
affecting the sources. Multi-epoch observation is thus used to modulate the
deflection ON (Sun between fields) and OFF (Sun at large distance).

The benefits to systematic error control of simultaneous observation of 2+2 or
4+4 fields are not only related to the \emph{increased efficiency} (same
exposure time on different objects multiplexed on the detector), but above all
to \emph{real time compensation of systematic errors}, affecting in common mode 
all stars, independently from the deflection modulation phase. 
The superposition can be achieved with techniques similar to those adopted in 
Hipparcos and Gaia, i.e. a \emph{beam combiner} (BC) folding onto the same 
detector the images of the two fields. The separation between observing 
directions, materialised by the BC, is the above mentioned \emph{base angle} 
(BA), with value $\sim 4^\circ$.

Most disturbance effects acting as systematic errors on individual star location
estimate, as background non-uniformity, nearby stars, proper motions, stellar
activity (e.g. pulsations), and chromaticity, are averaged out over the transit,
the stellar sample, and/or epochs throughout the mission.

\subsection{Estimation of the PPN $\beta$ parameter}
\label{subsec:beta}
The second main scientific goal of the mission is the estimation of the PPN
parameter $\beta$ to a few parts per million. The perihelion shift excess of the
orbits of massive bodies is proportional to the factor $2\beta-\gamma$ which
allows one to estimate the value of $\beta$ since $\gamma$ is known to a better
accuracy level. The sensitivity of this effect to higher eccentricities and
smaller orbits makes Mercury a favourite candidate for this kind of measurements
in the Solar System, and in fact radar observations of this planet accumulated
between 1966 and 1990, resulted in an accuracy of $3\cdot10^{-3}$ for the factor
$2\beta-\gamma$ \cite{1990grg..conf..313S}. 
The importance of an accurate estimation of this parameter, however, is 
easier to understand if one considers that it enters in the factor 
$4\beta-\gamma-3$ associated to the so-called Nordtvedt effect, whose 
existence would imply a violation of the Strong Equivalence Principle in 
the weak gravity regime of the Solar System. 
This was put to test by Williams et al. \cite{2004PhRvL..93z1101W} with 
LLR measurements, which reached the same $10^{-4}$ accuracy level obtained 
with the grand fits. 
The precise measurement of $\beta$ has also important cosmological 
implications, since in $f(R)$ theories it depends on the derivative 
of $\gamma$ with respect to the scalar of curvature 
$\mathrm{d}\gamma/\mathrm{d}R$ \cite{2005PhRvD..72d4022C}. 
A non-zero value of $\beta$ would therefore imply a different behaviour 
of gravity at different distance ranges, providing an alternative explanation 
of the observed acceleration of the expansion of the Universe at large scale, 
which is currently attributed to the existence of Dark Energy.

As pointed out in Capozziello and Tsujikawa \cite{2008PhRvD..77j7501C}, several
$f(R)$ models are able to reproduce the characteristics of a $\Lambda$CDM
scenario with Dark Energy, which cannot be ruled out by the present constraints
on the PPN parameters.

GAME is specifically designed to perform high-accuracy astrometric 
measurements in the vicinity of the Sun. 
This naturally suggests the possibility of using part of its observing time 
for the determination of the orbit of some convenient set of bodies of the 
Solar System, in particular inner planets. 

\begin{figure}
  \begin{center}
    \includegraphics[width=0.7\textwidth]{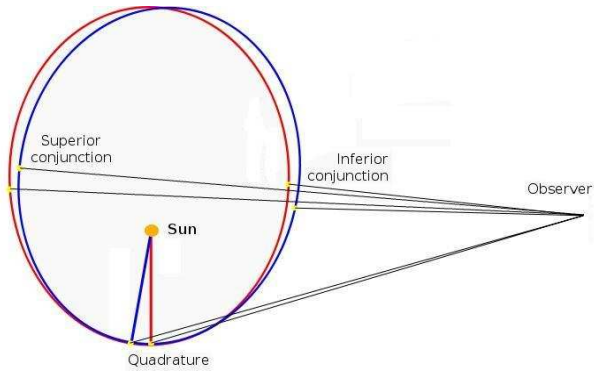}
  \end{center}
\caption{From a geometrical point of view, the perihelion shift 
excess can be observed with the highest SNR ratio superior conjunction, 
not accessible from the ground. 
The benefit is due to the brighter apparent magnitude of the planet.} 
\label{fig:Mercury}
\end{figure}

It is worth noting that from ground, using most existing telescopes, 
it is \emph{not} possible to observe Mercury close to 
conjunction, as in Fig.~\ref{fig:Mercury}, where its position determination 
is most favourable to evidence the precession effects. 

\subsection{Oblate and moving planets}
\label{subsec:quadrupole}
High precision astrometry around the major planets of the Solar System may
discriminate, through estimate of the light deflection, their mass monopole and
quadrupole effects, as well as gravito-dynamical contributions. Observations
close to Jupiter's limb (or other giant planets like Saturn) will provide the
estimation of the light deflection induced by the quadrupole component of its
gravitational field. The possibility of measuring this effect, foreseen by
Relativistic theories of gravity but still never addressed by any experiment,
has been simulated for a Gaia-like mission for the first time in Crosta and
Mignard \cite{2006CQGra..23.4853C}, and this concept is applicable to GAME. 
The magnitude of the contribution to the total light deflection coming from 
the quadrupole component of the gravitational field is 
$\left|\delta\psi_\mathrm{Q}\right| = 
2\left(1+\gamma\right)MJ_2\left(R^2/b^3\right) 
f\left(\theta,\psi\right) 
\varepsilon$,
where $b$ is the impact parameter with respect to the deflecting body, 
$\theta$ is the latitude of the observed star relative to the 
equatorial plane of the deflecting object, $\psi$ is the angular 
distance from the same object, 
and $\varepsilon$ is the parameter to be estimated, whose value is 1 in GR. 
Typically, since the effect is proportional to $b^{-3}$ , one can skip the
dependence from $\psi$ since at large angular distance the effect is too small.
From this formula, it is easy to deduce that
$\sigma_{\delta\psi_\mathrm{Q}}/\delta\psi_\mathrm{Q} = 
\sigma_\varepsilon/\varepsilon$.

Considering reasonable average stellar densities for targets brighter than
$V=9$, and a FOV amplitude of $\sim30'$, one can expect to estimate the
parameter $\epsilon$ with a final relative accuracy in the range
$10^{-3}-10^{-4}$ using about $10\%$ of the 5 years mission lifetime.

\subsection{Extra-solar planets}
\label{subsec:exoplanets}
The high-precision fully differential astrometry achievable on bright targets
and its high-accuracy, high-cadence photometry make GAME an ideal instrument to
undertake a multi-tiered program of targeted measurements of well-selected,
known extra-solar planetary systems, with the overarching goal of critically
deepening our understanding of key issues in exoplanet astrophysics that are
likely to have significant margins of improvement in a decade from now. We will
devise a Flexible Observing Strategy (FOS) to implement a three-fold exoplanet
science program focused on:
\begin{enumerate}
\item astrometry of exoplanets from long-term high-precision RV monitoring;
\item follow-up studies of Gaia-detected exo-planets;
\item astro-photometric investigations of transiting planetary systems.
\end{enumerate}

The improvement of our knowledge of the complex processes of planet formation
and evolution as a function of stellar characteristics (mass, metallicity, age)
and system's architecture can only be achieved through their long-term,
multi-technique, multi-wavelength monitoring, with the four-fold goal of
\begin{enumerate}
\item\label{g1} improving the orbital phase sampling;
\item\label{g2} looking for planetary companions at all orbital periods;
\item\label{g3} looking for low-/high-mass components (Super Earths
to Jupiters);
\item\label{g4} refining the characterization of multiple planet systems
(including accurate coplanarity measurements).
\end{enumerate}

GAME's fully differential astrometric approach will allow us to precisely
measure the targets' position against a set of 6 to 8 reference stars with
magnitude $V<11$~mag. These will be selected to be distant ($D>1$~kpc) $G$-$K$
giants with no close-in stellar companions (the astrometric motion induced by
planetary mass companions being negligible at such distances, given GAME's
reference single-measurement precision).
\\ 
Robust identification of distant giants will be readily carried out exploiting
Gaia's direct parallax measurements, with careful selection of the targets and
their references.
\\ 
About $30\%$ of the GAME observing time may be devoted to exoplanet
science.

\subsection{Near Earth  objects and Solar System dynamics }
\label{subsec:neo}
The combination of observation on or near the Ecliptic, and of high cadence
window readout, provides GAME with an unmatched capability for astrometric
measurement of a large set of solar system objects, from major planets to main
belt asteroids, on a common precise reference frame in the visible spectral
range. 
The contribution to the ``grand-fits'' of Solar System dynamics 
\cite{2009A&A...507.1675F}, adding to and bridging the range of available 
visible and radar observations, will significantly improve the precision, 
and above all self-consistency, of the estimate of fundamental parameters 
and individual ephemeris.
\\ 
The prolonged observation of the inner Solar System will allow precise orbital
parameter determination for many known Near Earth Objects, apart providing the
option for detection of new ones, thus helping in timely evaluation of potential
hazards to our planet.
\\ 
Additional targets of opportunity may arise, as detection of Sun-grazing comets
and observation of close encounters between asteroids.

\section{Mission profile and instrument concept }
\label{sec:profile}
GAME requires high astrometric precision on four (``Sun-ward'') fields close to
the Sun to detect their deflection; a large field of view is required to achieve
a large star sample. Observation in low deflection conditions is performed on
four (``Out-ward'') fields in the opposite direction, simultaneously
(Fig.~\ref{fig:ObsGeometry}).

The simultaneous observation of multiple fields, both close to the Sun and in
the opposite direction, is adopted not only to increase the observations, but
above all to minimize systematic errors by differential measurement, thus
alleviating the engineering requirements on payload and spacecraft.

GAME is based on a single main instrument, a 1.5~m diameter telescope, with
coronagraphic and multiple field beam combination sub-systems. It is compatible,
in terms of mass and volume budgets, with additional payloads.

The satellite is oriented with a side always set toward the Sun (within
$35^\circ$), thus simplifying the design of payload thermal control and solar
panel allocation; the detector radiator is naturally set on the dark side.

The telescope is endowed with a pupil mask and a folding mirror for injection of
the out-ward field beams. Each sub-aperture acts as a coronagraph for rejection
of the Sun disk at the $R\leq10^{-8}$ level, and of the inner Corona, whereas,
with respect to the stellar fields, the set of sub-pupils works as a Fizeau
interferometer, feeding the underlying monolithic telescope. The background is
fundamentally limited by the Corona at $2^\circ$ from the Sun centre, to
$\sim9$~mag per square arcsec. Suppression of the Sun disk light requires
adoption of coronagraphic techniques in the instrument design.

The payload technology is mostly based on Gaia inheritance for astrometry; the
coronagraphic sub-system adopts the concept of inverted occulter from e.g.\ the
METIS instrument on the Solar Orbiter. The beam combiner folds the four front
and four rear fields over the common focal plane.

GAME is compatible with a launch on Soyuz/Fregat from the Guiana Space Center,
with injection into the selected Sun-synchronous orbit, with height of
$\sim1500$~km, in order to ensure observing time close to $100\%$ and to retain
a stable power and thermal environment to the satellite.

A single ground station, with average visibility of order of 1~hour/day and a
data rate of order of 8~Mb/s, is considered sufficient to the mission needs. The
data storage and processing requirements are not challenging.


\begin{figure}
  \begin{center}
    \includegraphics[width=0.49\textwidth]{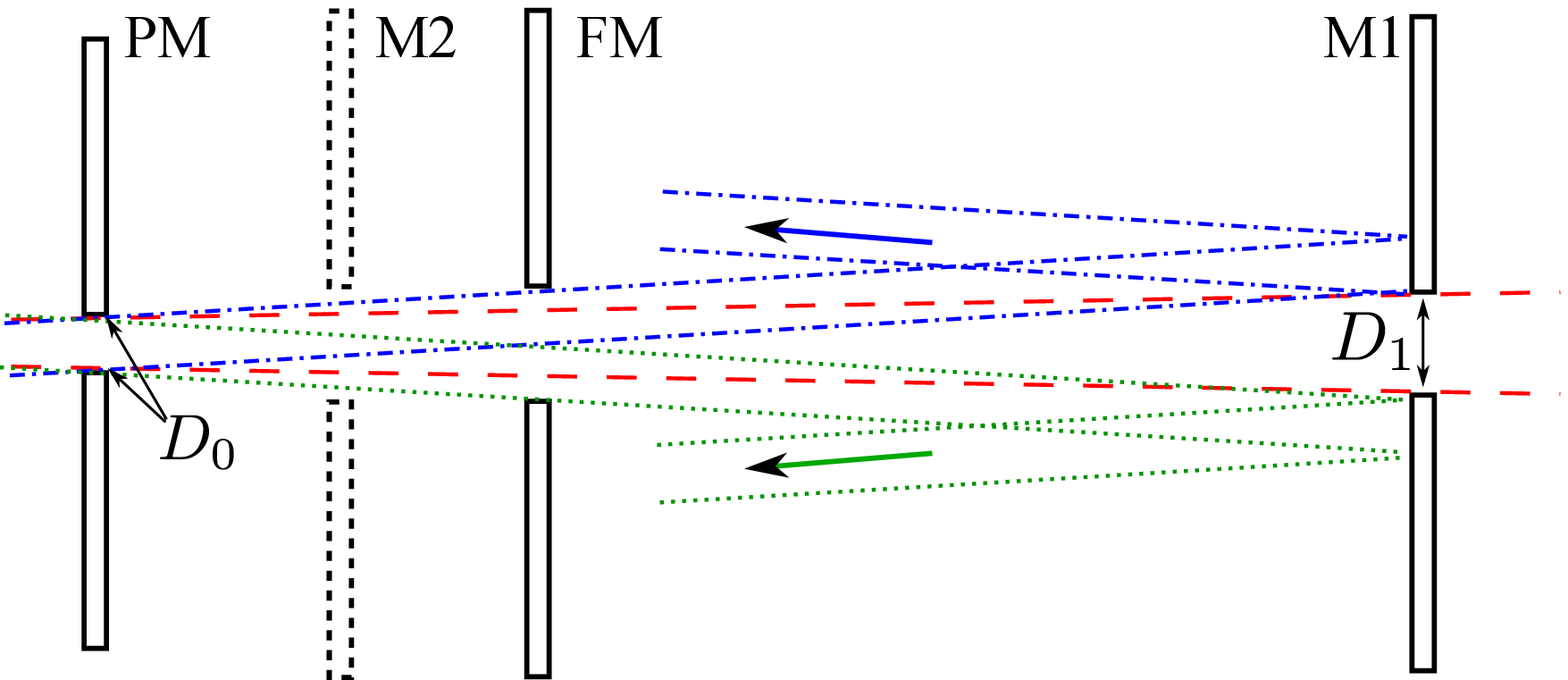}
    \includegraphics[width=0.49\textwidth]{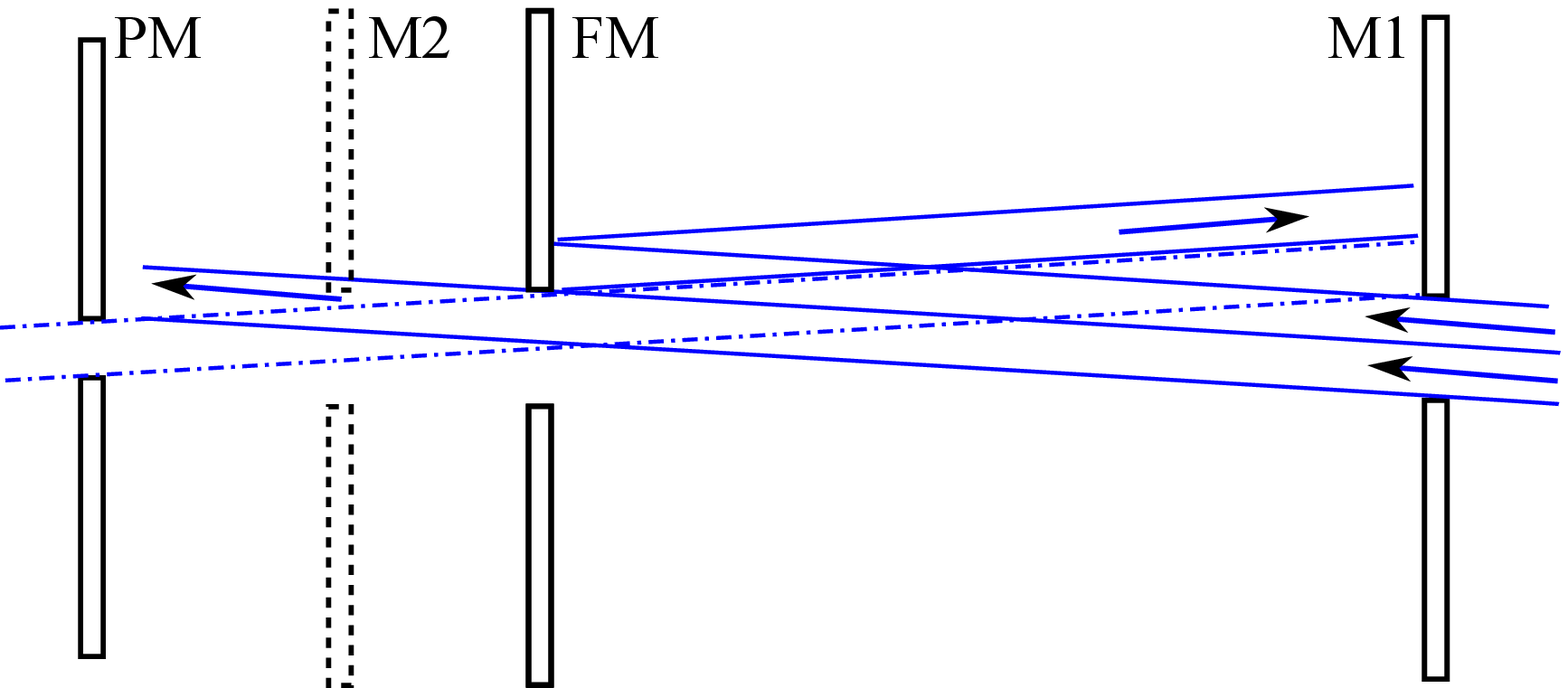}
  \end{center}
\caption{Rejection of one Sun beam and injection of the two related 
Sun-ward field beams (left), 
entering the system from the left through the pupil mask (PM), and reaching the primary mirror (M1) through the apertures on the folding mirror (FM). 
The Sun beam (dashed line) is dumped to space through the M1 apertures. 
The stellar beams (dotted and dash-dot lines) reach the reflecting surface 
of M1 and are focused toward the detector. 
Injection of one outward beam (right, solid line.) from the apertures on M1, 
by reflection on FM. It is then propagated back toward M1, parallel to 
corresponding Sun-ward beams (dash-dot lines). 
The secondary mirror M2 (dashed line rectangle) is between PM and FM. 
}
\label{fig:BeamInject}
\end{figure}

\subsection{The measurement technique}
\label{subsec:measurement}
GAME observes in step-and-stare mode four fields around the Sun, at radial
distance $2^\circ$ (deflection: $\sim0''\!\!.233$ ) from the Sun
centre, and set at $90^\circ$ from each other. The nominal field placement is
along the ecliptic longitude and latitude axes, labelled respectively North,
South, East and West (N, S, E, W).

In addition to the four ``Sun-ward'' fields, GAME also observes at the same
time, and using as much as possible the same parts of the instrument, four
``outward'' fields with the same relative placement in the direction opposite to
the Sun. Re-observations of the same fields in different epochs may be adopted
to improve on calibration issues, requiring about $50\%$ of the observing time
for the main science goal of GAME. The remaining time can be devoted to
additional science and contingency.

Subsequent exposures of the superposed fields are taken, to compute the
photo-centre location of each star image. Each photo-centre displacement between
deflection ON and OFF epochs provides an independent estimate of the angular
value of light deflection, which can then be cumulated over the star sample.
Relaxed requirements are imposed on a priori knowledge of star parameters and on
pointing accuracy; accurate reconstruction of attitude and astrometry is
expected as a side product of data processing.

The star displacement due to light deflection must be evaluated with respect to
a reliable reference frame. The simultaneous observation of two fields in
appropriate positions, e.g. at the same radial distance to the Sun centre, in
opposite directions, provide simple detection by relative astrometry, since the
stars are affected by deflection with similar amplitude and opposite sign along
the line joining the fields and the Sun centre.

The high density strip of intersection of the Ecliptic and Galactic plane has
$\sim30^\circ$ length, and is spanned in about 1 month by the apparent motion of
the Sun associated to Earth's orbital motion in about one month.

\subsection{Instrument conceptual design and key characteristics}
\label{subsec:instrument}
The GAME optical concept is based on Fizeau interferometry, to achieve a
convenient trade-off between the angular resolution needed for precision
astrometry, and coronagraphic requirements, applied to small apertures achieved
by pupil masking on the underlying telescope. Below the beam path related to a
single input aperture is described; afterwards, the combination of several
individual apertures is considered.

\begin{figure}
  \begin{center}
    {}\hfill\includegraphics[width=0.39\textwidth]{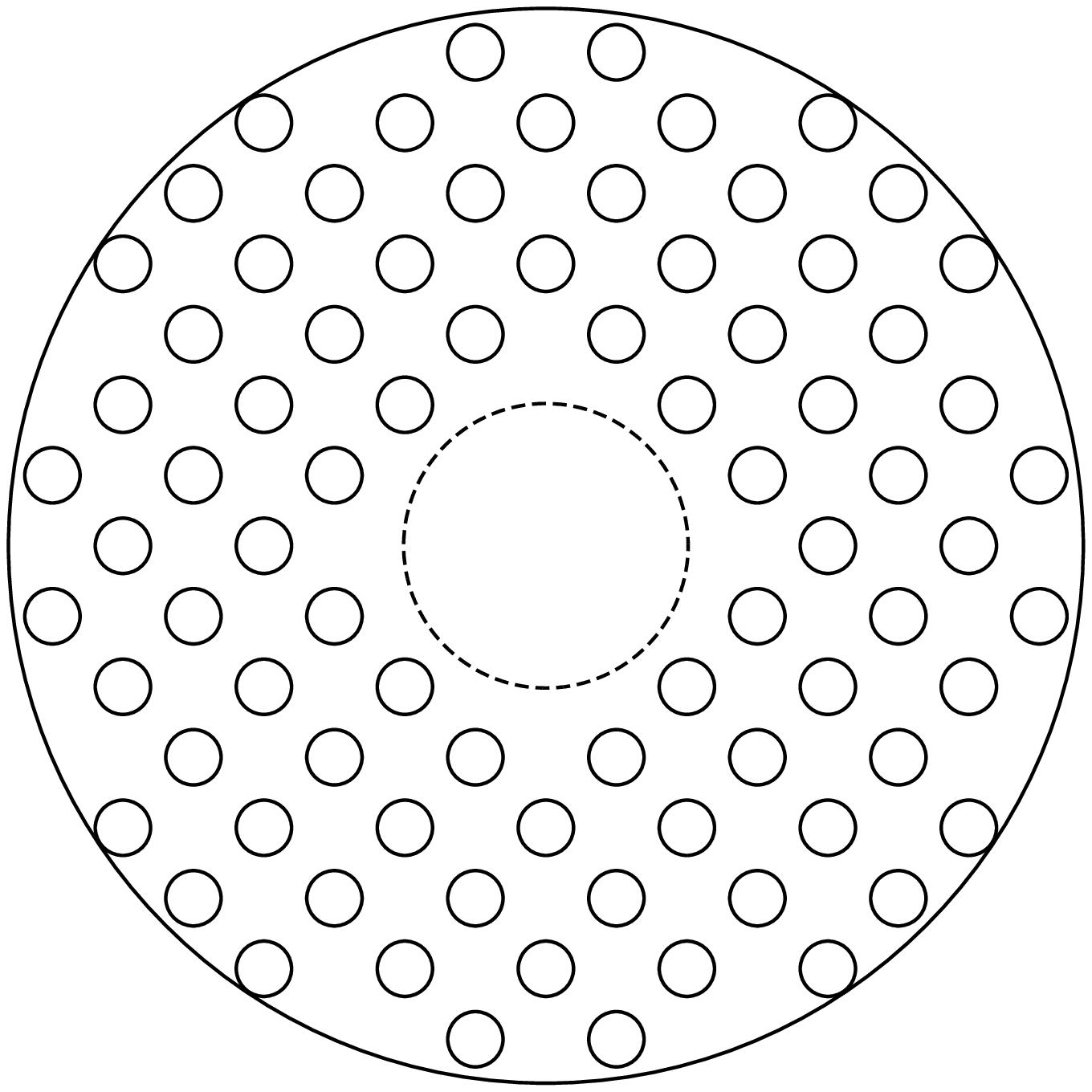}
    \hfill\includegraphics[width=0.39\textwidth]{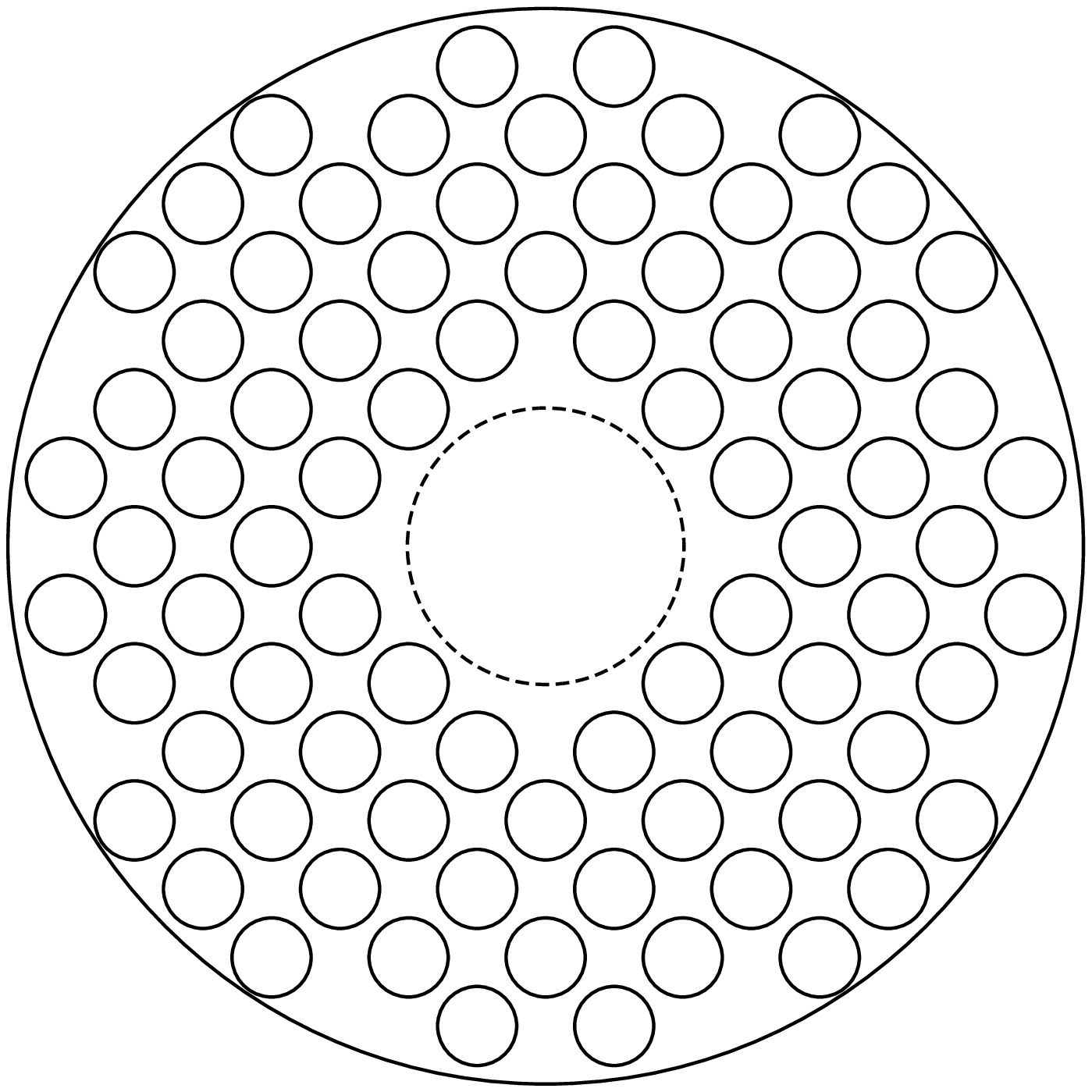}\hfill
    \includegraphics[width=0.2\textwidth]{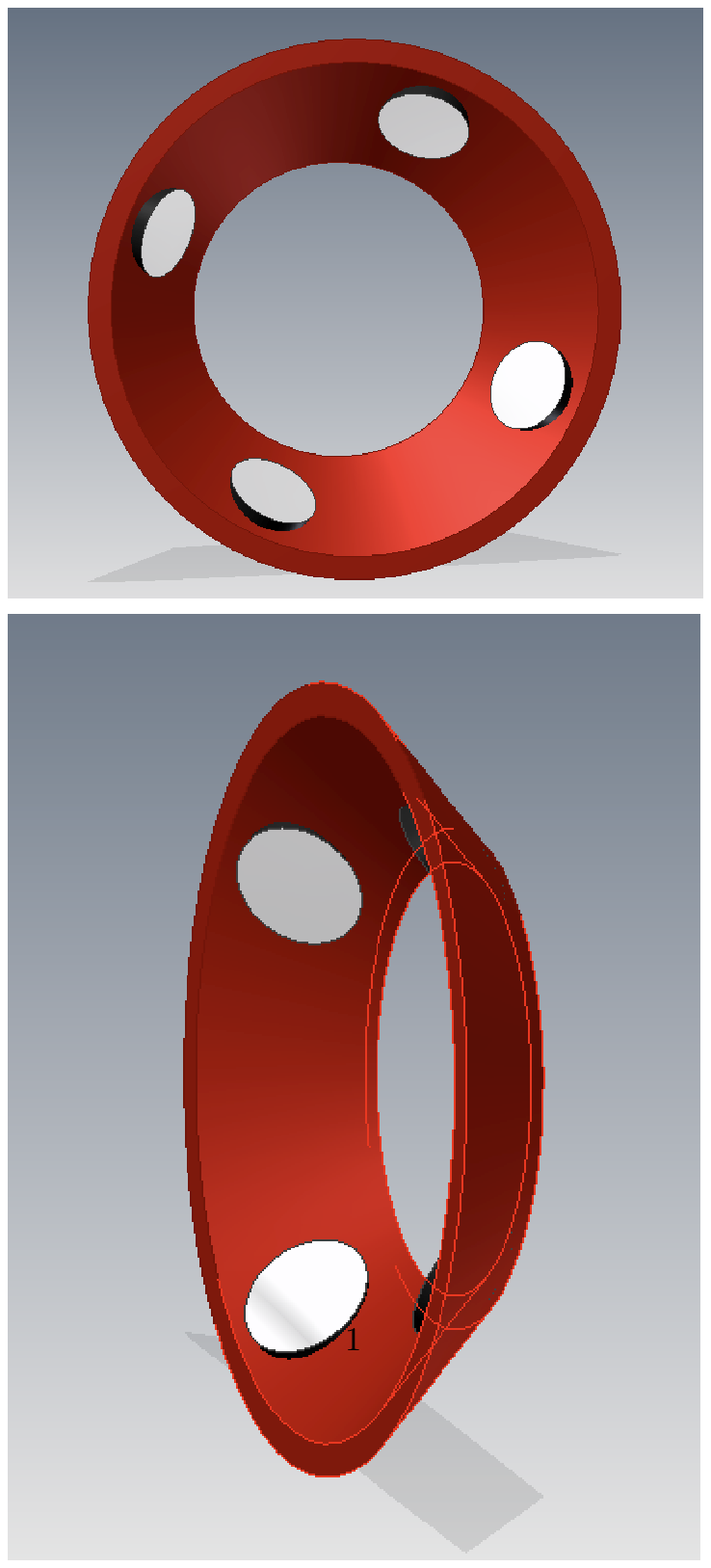}
  \end{center}
\caption{Fizeau mask on the Pupil Mask PM (left) and output apertures on Primary
mirror M1 (center); the Beam Combiner concept (right).}
\label{fig:PupilBC}
\end{figure}

The elementary aperture on the pupil mask PM is circular, with diameter
$D_0\simeq70$~mm. It is matched to an output aperture on the primary mirror M1
with diameter $D_1\simeq105$~mm, so that the mirror surface is not directly
faced to the Sun. Aperture centres are aligned in the Sun direction, and
$D_1>D_0$ to account for Sun diameter and diffraction margins. Diameters and
separations are set to avoid vignetting on the primary mirror of the stellar
beams from the observed fields. Apertures are endowed with serrated edges to
reduce the diffraction of the Sun beam within the instrument. A schematic view
of the basic layout is shown in the left side of Fig.~\ref{fig:BeamInject}. The Sun
beams from PM (dashed lines) are sent, through the M1 holes, to outer space,
without the need for e.g. photon traps. The stellar beams from two Sun-ward
fields (e.g., N and S), shown as dotted and dash-dot lines, are separated from
Sun beams by geometric optics on M1. The additional Sun-ward beams (E and W) are
orthogonal to the drawing plane.

The apertures on M1 also allow photons from star in the outward fields to get
into the system, as shown in the right side of Fig.~\ref{fig:BeamInject}. 
The folding
mirror FM between PM and M1 is used to inject the out-ward beams back on M1 and
into the telescope. Thus, the out-ward fields are imaged on the focal plane,
superposed to the Sun-ward fields, and using mostly the same optical system,
with the only addition of the flat mirror FM, acting as a ``vanity mirror'' for
the telescope. The outward beams are partially vignetted by the apertures on
both FM and M1, but, thanks to the larger diameter, still provide a photon
budget comparable to that of the Sun-ward fields. The elementary aperture
geometry introduces a marginal modulation of the image profile.

The aperture geometry is replicated several times in two dimensions, thus
providing the desired Fizeau mask on PM, generating images with resolution
comparable with the full underlying telescope for each of the observed fields.
The desired \emph{four field instrument, pair-wise symmetric and using as far as
possible the same components in common mode\/}, is thus achieved.

The effective diameter of the pupil array on PM is 1.44~m, corresponding to the
observing resolution; the overall diameter of M1 is 1.55~m, larger because of
the beam divergence among fields. A representation of the aperture array on PM
and M1 is shown in Fig.~\ref{fig:PupilBC}. The pierced mirror is a potentially
critical aspect of the mission with respect to manufacturing / stability.

\begin{figure}
  \begin{center}
    \scalebox{0.6}{\includegraphics*[0.5cm,11.5cm][20cm,18cm]{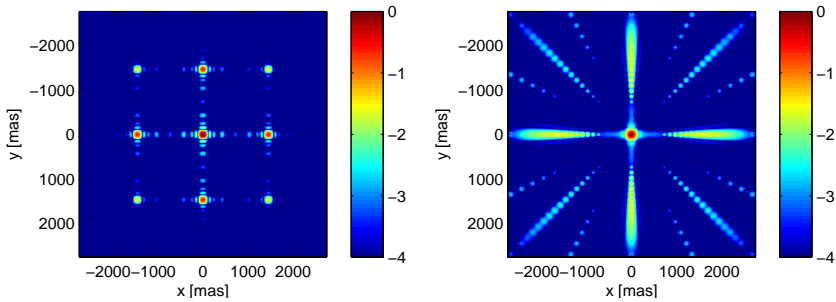}}
   \end{center}
\caption{Monochromatic PSF, $\lambda=600$~nm (left); polychromatic PSF for a G2V
star (right); logarithmic units are selected for the intensity.}
\label{fig:PSF}
\end{figure}

The imaging performance of the Fizeau array is shown in Fig.~\ref{fig:PSF}, in
logarithmic units, respectively for the monochromatic (left) and polychromatic (right) PSF.
The central lobe size is comparable to the full aperture PSF of the underlying
aperture ($\mathrm{FWHM}<100$~mas), and the regular structure in the side lobes
is significantly smeared out by the source spectral distribution.

The detector is a mosaic of 72 EEV CCD42-80, 2k×4k, $13.5~\mu$m pixels. The
pixel size and sampling requirements on the diffraction limited image impose an
effective focal length $EFL=40$~m. The angular pixel size is then respectively
$70$~mas. Window readout around the targets of interest is performed, to
alleviate the electronics, data storage and telemetry requirements. A detection
/ confirmation function (similar to Gaia's) can be implemented on the leading
edge of the detector. Exposures on most stars have duration 2 to 3 minutes,
whereas CCDs imaging a few bright stars are read at faster rate, to prevent
saturation, and in support to on-board attitude reconstruction.

\begin{figure}
  \begin{center}
\includegraphics[trim=1cm 0 0 0,clip,width=0.7\textwidth,angle=-90]{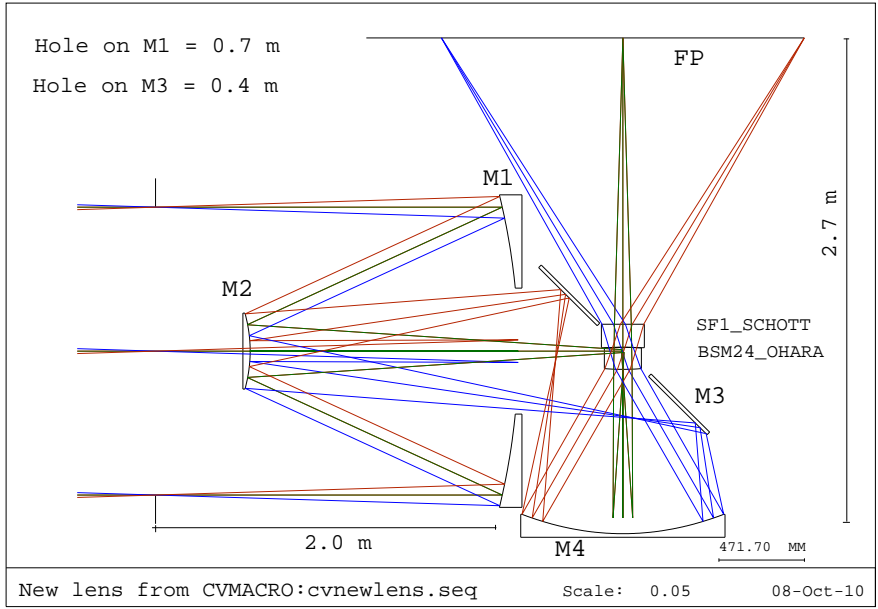}
  \end{center}
\caption{ Unfolded optical design of the GAME telescope. The pupil mask PM 
(at 2~m from the primary mirror) and folding mirror FM (close to the 
secondary mirror M2) are not shown to evidence the telescope proper. 
The beam combiner, replacing the tertiary mirror M3, superposes on the focal plane the fields of view, shown separated in figure; this also reduces the 
required size of the fourth mirror M4. }
\label{fig:Telescope}
\end{figure}

An optical design compliant with the GAME requirements is shown in
Fig.~\ref{fig:Telescope}, avoiding for clarity the folding by the beam combiner and
other mirrors into the satellite envelope. The on-axis configuration is based on
a classical Ritchey-Chr\'etien scheme for the primary (M1) and secondary (M2).
The optical performance is good (RMS $\mathrm{WFE}<60$~nm) over a $24'$ diameter
field. The folding mirror FM has diameter comparable to M1, i.e. 1.55~m; its alignment
has quite relaxed requirements.

The crucial element multiplexing the four off-axis directions (fields N, S, E,
W) at the telescope input onto the same optical axis of the following optics is
the beam combiner (BC), taking the place of the M3 mirror in 
Fig.~\ref{fig:Telescope}. The 
BC, Fig.~\ref{fig:PupilBC} on the right, is composed of four individual mirrors set
on a high stability supporting structure, with $90^\circ$ rotational symmetry of
placement. Each mirror introduces Korsch-like correction for its associated
field.

\subsection{Performance assessment}
\label{subsec:performance}
The measurement scheme of GAME is \emph{fully differential\/}, since it is based
on determination of the variation between epochs of the angular distance between
stars in selected fields. By design, the instrument is as symmetric as possible
with respect to the fields of view, and stable between epochs, also imposing a
stable observing geometry with respect to the Sun and Earth, the main sources of
thermal disturbances.

The measurement noise performance (random error) depends on the location
precision on individual sources, and on the total number of sources in either
field, therefore on the angular size of the field and on the number of different
fields observed. The individual location precision depends on the instrument
characteristics, source spectral type and magnitude, and on background level
\cite{1998PASP..110..848G,2001A&A...367..362G}.

\begin{figure}
  \begin{center}
    \includegraphics[width=0.6\textwidth]{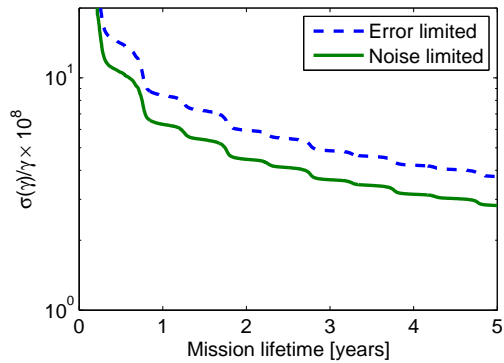}
  \end{center}
\caption{ Performance on PPN $\gamma$ vs. mission duration.}
\label{fig:PerfGamma}
\end{figure}

The noise limit on the PPN $\gamma$ estimate includes an overall degradation of
the diffraction limited imaging performance by $20\%$, considering realistic
optical response, pointing jitter, and other random contributions. 
The performance of GAME on $\gamma$ is shown in Fig.~\ref{fig:PerfGamma}, 
in terms of noise limit vs. the mission lifetime (solid line). 
The error limit (dashed line) includes a further $30\%$ degradation associated 
to uncalibrated systematic errors, degrading the realistic performance to 
$\sigma(\gamma)/\gamma = 3.8\times10^{-8}$. 

The performance on $\beta$ is preliminarily estimated in the framework 
described in Sec.~\ref{subsec:beta}, by considering absolute measurements 
of the position of Mercury close to the superior conjunction, and assuming 
the motion of all other planets known with adequate precision. 
The calculations suggest that the accuracy level of 
$\sigma(\beta)/\beta = 2.5\cdot10^{-6}$, 
i.e. two orders of magnitude better than the present experimental results, 
could be reached with some $5\cdot10^{8}$ elementary observations, 
requiring a total of just about 10 days over the 5 years mission lifetime. 

\section{Additional system requirements}
\label{sec:requirements}
GAME, similarly to Gaia, has a single main payload, strictly integrated in the
satellite structure. The proposed procurement approach is therefore similar,
assuming that the satellite, payload and operation are fully under ESA
responsibility. The payload fits within an envelope of diameter $D=3.4$~m, with
height $H=1.8$~m. The service module has height $H=1.5$~m, and the payload mass is
below 840~kg, including $30\%$ contingency.

The peculiar aspect of GAME concerns observation close to the Sun, during its
transit over the selected high stellar density regions, in order to detect the
largest possible value of light deflection. Step-and-stare pointing is required.
The pointing jitter must be limited to avoid image degradation by more than a
fraction of a pixel, so that the requirement is set to 15 mas RMS over 120~s
standard exposures.

A simple computation of the basic requirements can refer to the density in the
most populated regions: $\sim130$ stars / CCD. Thus, the full detector data flow
is $\sim3$~GB/day. The detector operation is based on windowed readout, with
120~s exposures. Stars brighter than $V\simeq13$~mag require that the associated
CCD is operated at a faster rate to prevent saturation. Short exposure time is
also required for observation of Solar System planets and some additional
science cases involving bright stars ($<10$~mag).

The GAME concept builds upon inheritance of basic principles and technology from
space missions already flown or in an advanced state of development, both on the
astrometric side (resp. Hipparcos and Gaia) and on the proven concepts of
external occulter, apodised apertures and internal Lyot stop (SoHO, SCORE, METIS
for the Solar Orbiter).

The telescope technology mostly comes from Gaia, with intensive adoption of
common mode rejection techniques for external disturbances to ensure the
achievement of the mission goals in the operating environment.

The GAME large mirror (1.5~m) can be manufactured in silicon carbide with
existing facilities. The main criticality concerns the manufacturing of large
mirrors endowed with an array of apertures to implement the desired
coronagraphic Fizeau interferometer.

In the main science mode the spacecraft will be pointing at the Sun and perform
simultaneous exposures with integration time of $\sim120$~s on all selected
targets in the four + four superposed FOVs. The pointing stability required is
of the order of 10~mas RMS over the 120~s integration time.

For the additional science observations, GAME must be able to slew to different
pointing positions, within $\sim35^\circ$ to the Sun, within a reasonable time,
e.g. to point a target in 10 minutes.

The data handling is limited to the selected observation windows on the CCD. The
average telemetry rate required is 300~kbit/s, assuming sustained data transfer
over 24~hours, corresponding to about 7~Mbit/s over a 1~hr visibility window. An
on board memory buffer is foreseen, corresponding to 24-48h of normal operation
data flow (24-48~Gbit=3-6~GB).

The computing power requirement is limited, since the baseline is to send most
of the unprocessed data to ground. Only 10-20 bright stars are read at a faster
rate (1-30~s exposure, usually in different CCDs), for use by the on-board
attitude estimate algorithms in support to AOCS.

\section{Conclusions}
The smart combination of modern astrometric and coronagraphic techniques 
allows the definition of a mission concept able to provide unprecedented 
results on Fundamental Physics by estimation of the PPN parameters 
$\gamma$ and $\beta$, respectively to the $10^{-8}$ and $10^{-6}$ range. 
Also, a number of additional astrophysical topics may be addressed by 
the high precision astrometry and high cadence photometry capabilities 
of GAME. 

%

\begin{acknowledgements}
We are grateful to the colleagues, from our and other Institutes, who 
contributed to the GAME concept development throughout the past few 
years, and recently to the implementation of a proposal for a medium 
mission in response to the M3 ESA call. 
The activity was partially supported by the Italian Space Agency 
contract I/058/10/0. 
\end{acknowledgements}

\bibliographystyle{spmpsci}      


\end{document}